\documentclass
[twocolumn,preprintnumbers,amsmath,amssymb,superscriptaddress,pra]{revtex4}

\usepackage{epsfig}
\usepackage{amsmath}
\usepackage{graphicx}
\usepackage{bm}
\usepackage{ulem}

\usepackage{wasysym} %for 'approximately greater'-symbol

\usepackage[dvips]{color}

\begin{document}

\title{Monocrystalline Al$_{x}$Ga$_{1-x}$As heterostructures for high-reflectivity high-Q micromechanical resonators in the megahertz regime}

\author{Garrett D. Cole}
\email{cole35@llnl.gov} \affiliation{Center for Micro- and Nanotechnologies, Lawrence Livermore National Laboratory, 7000 East Avenue, Livermore, CA 94550, USA}

\author{Simon Gr\"oblacher}
\affiliation{Institute for Quantum Optics and Quantum Information (IQOQI), Austrian Academy of Sciences, Boltzmanngasse 3, A--1090 Vienna, Austria}

\author{Katharina Gugler}
\affiliation{Institute for Quantum Optics and Quantum Information (IQOQI), Austrian Academy of Sciences, Boltzmanngasse 3, A--1090 Vienna, Austria}

\author{Sylvain Gigan
\footnote{Permanent Address: Laboratoire Photon et Mati\`{e}re, Ecole Superieure de Physique et de Chimie Industrielle, CNRS-UPR A0005, 10 rue Vauquelin, 75005 Paris, France} \affiliation{Institute for Quantum Optics and Quantum Information (IQOQI), Austrian Academy of Sciences, Boltzmanngasse 3, A--1090 Vienna, Austria}
}

\author{Markus Aspelmeyer}
\affiliation{Institute for Quantum Optics and Quantum Information (IQOQI), Austrian Academy of Sciences, Boltzmanngasse 3, A--1090 Vienna, Austria}

\begin{abstract}
We present high-performance megahertz micromechanical oscillators based on freestanding epitaxial Al$_{x}$Ga$_{1-x}$As distributed Bragg reflectors. Compared with dielectric reflectors, the low mechanical loss of the monocrystalline heterostructure gives rise to significant improvements in the achievable mechanical quality factor Q while simultaneously exhibiting near unity reflectivity. Experimental characterization yields an optical reflectivity exceeding 99.98\% and mechanical quality factors up to 20 000 at 4~K. This materials system is not only an interesting candidate for optical coatings with ultralow thermal noise, but also provides a promising path towards quantum optical control of massive micromechanical mirrors~\cite{Cole2008}.
\end{abstract}

\maketitle

\newpage

High-quality Bragg mirrors with small mechanical dissipation have generated recent interest due to their versatile use in both fundamental and applied sciences. Specifically, mechanical dissipation  in optical coatings is known to limit the performance of high-finesse cavity applications, in particular gravitational wave interferometry~\cite{Harry2006} and laser frequency stabilization for optical clocks~\cite{Numata2004}, because of residual phase noise, also referred to as coating thermal noise~\cite{Harry2002}. On the other hand, microstructures of high mechanical and optical quality have become a leading candidate to achieve quantum optical control of mechanical systems. One specific goal in this emerging field of quantum optomechanics is to combine the concepts of cavity quantum optics with radiation-pressure coupling to generate and detect quantum states of massive mechanical systems such as the quantum ground state~\cite{Wilson-Rae2007,Marquardt2007,Genes2007} or even entangled quantum states~\cite{Bose1997,Zhang2003,Vitali2007}. The recent demonstrations of cavity-assisted laser-cooling of mechanical modes~\cite{Metzger2004,Gigan2006,Arcizet2006b,Kippenberg2007} can be considered an important milestone in this direction.

Most of these schemes rely crucially on mechanical structures that combine both high optical reflectivity R and low mechanical dissipation, i.e. a high quality factor Q of the mechanical mode of interest. In addition, entering the quantum regime will require operation in the so-called sideband-limited regime~\cite{Wilson-Rae2007,Marquardt2007,Genes2007}, in which the cavity bandwidth of the optomechanical device is much smaller than the mechanical resonance frequency. While toroidal microcavities have recently shown such performance~\cite{Schliesser2008}, high-quality distributed Bragg reflectors (DBRs) in combination with Fabry-P\'{e}rot cavities have not yet reached this regime~\cite{Boehm2006,Gigan2006,Arcizet2006b,Kleckner2006b}. For example, whereas DBRs based on SiO$_{2}$/Ta$_{2}$O$_{5}$ can achieve R values in excess of 99.99\%~\cite{Rempe1992}, the mechanical quality factor of free-standing DBRs is limited to below 3000 due to internal losses in the Ta$_{2}$O$_{5}$ layers~\cite{Groeblacher2008a}. It is interesting to note that the low Q-value obtained with these devices is consistent with the coating loss angles observed in the LIGO studies of gravitational wave detector coatings of the same material~\cite{Harry2006,Harry2002}. On the other hand, the use of SiO$_{2}$/TiO$_{2}$-based DBRs has led to the demonstration of mechanical quality factors approaching 10 000 at room temperature~\cite{Gigan2006}; there, however, optical absorption in TiO$_{2}$ at 1064~nm both limits the reflectivity and results in residual photothermal effects.

The concept outlined here seeks to improve upon these previous works by fabricating the oscillator directly from a \textit{single-crystal} Bragg reflector. In particular, the use of compound semiconductor materials such as GaAs and related alloys allows for the generation of arbitrary stacks of high-index-contrast materials, resulting in significant improvements in the achievable mechanical quality factor. Given the alleviation of the dangling bonds typically found in amorphous dielectric materials such as Ta$_{2}$O$_{5}$~\cite{Harry2002}, the use of a single-crystal mirror stack should allow for a significant reduction in the intrinsic damping, while maintaining excellent reflectivity. Neglecting support loss or modal coupling, mechanical dissipation in a single-crystal is ultimately limited by intrinsic processes such as thermoelastic damping, as well as phonon-phonon and phonon-electron interactions. Our devices do not approach this fundamental value but are most likely limited by extrensic effects including process-induced damage (e.g.,\ ion bombardment and surface roughness created during microfabrication) as well as acoustic loss to the surrounding support structure. For example, if thermoelastic damping were the lower limit to the mechanical dissipation of the device, we would expect a room temperature Q value of approximately $4\times10^{8}$ for a GaAs resonator~\cite{Braginsky1985}.

\begin{figure}[htbp]
\centerline{\includegraphics[width=.5\textwidth]{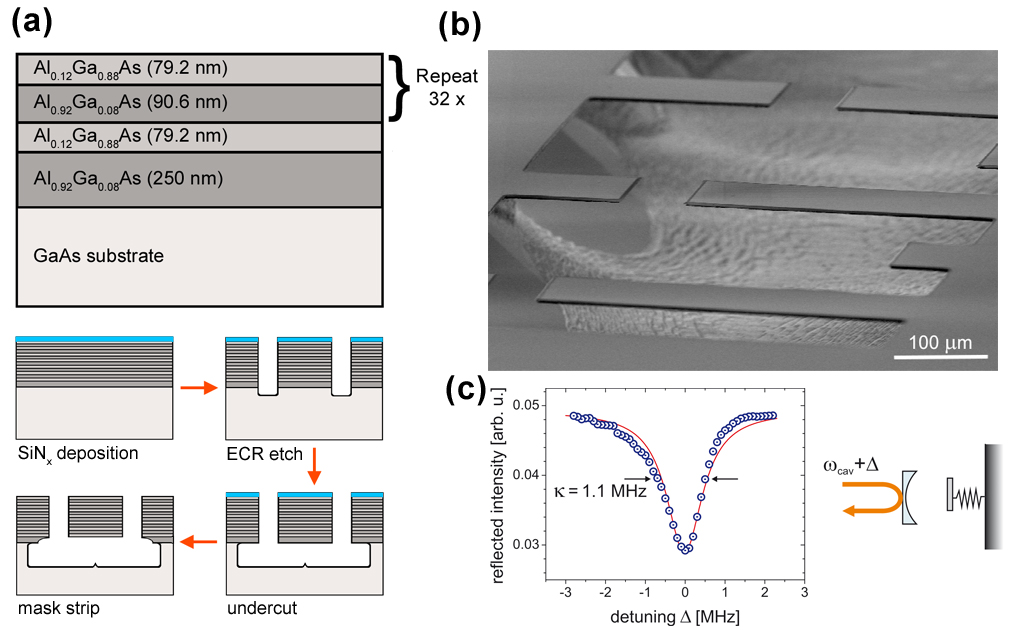}}
\caption{(a) Sketch of the initial layers constituting the Bragg mirror and illustration of the etch process used to fabricate free standing structures. (b) Micrograph of a group of cantilevers. The beams shown have a width of 50~$\mu$m and vary in length between 50 and 200~$\mu$m. (c) The finesse of the cavity is obtained by measuring the light reflected from the cavity as a function of laser detuning $\Delta$. The observed linewidth $\kappa$ of 1.1 MHz corresponds to an optical finesse of 5500.}
\end{figure}

Although a somewhat uncommon materials system for the development of micromechanical structures, GaAs and its alloys exhibit a number of advantageous properties~\cite{Hjort1993}. The direct bandgap optical transition in GaAs allows for the integration of optoelectronic functionality with micromechanical elements~\cite{Ukita1993}. Furthermore, the noncentrosymmetric nature of the zinc-blende crystal structure gives rise to an appreciable piezoelectric coefficient, allowing for efficient actuation or transduction in these materials. For our purposes, we take advantage of the ability to produce high-quality single-crystal Bragg stacks through the use of lattice-matched ternary alloys based on Al$_{x}$Ga$_{1-x}$As. These materials may be epitaxially grown as monocrystalline heterostructures via deposition methods such as molecular beam epitaxy (MBE) and metal-organic chemical vapor deposition. The ability to control the lattice matching condition through the use of alloying gives one the ability to "strain engineer" films in order to create built-in tensile or compressive stresses. In addition, variations in the aluminum composition allow for a wide range of selective etch chemistries over GaAs. Generally, these films display extremely high etch selectivites--in fact HF etching of the lattice-matched binary material AlAs versus GaAs exhibits a selectivity approaching $10^{7}:1$~\cite{Yablonovitch1987}. Al$_{x}$Ga$_{1-x}$As heterostructures may thus be processed using standard micromachining techniques to yield atomically flat optical surfaces that are ideal for optomechanical structures, as previously demonstrated in micromechanically-tunable surface-normal photonic devices~\cite{Cole2005,Maute2006,Huang2007}.

As shown in Fig.\ 1, the epitaxial materials structure for the monocrystalline oscillators consists of 32.5 periods of alternating Al$_{0.12}$Ga$_{0.88}$As (high index) and Al$_{0.92}$Ga$_{0.08}$As (low index), followed by a 250-nm thick high-aluminum content etch-protection layer, grown on a 3 in.\ semi-insulating GaAs substrate via MBE. In this design, the thick high-aluminum-content layer below the Bragg stack is included to protect the bottom of the mirror structure in subsequent processing steps. The peak reflectivity of the DBR is designed to be at 1078~nm at room temperature; in this case, the wavelength of maximum reflectivity is red-shifted to allow for thermo-optic effects upon cooling. The refractive index of the ternary compounds at cryogenic temperatures is estimated using the modified Afromowitz model developed in~\cite{Akulova1998}. Assuming no absorption and atomically smooth interfaces, the maximum reflectivity (after stripping the protective Al$_{0.92}$Ga$_{0.08}$As layer and with air cladding top and bottom) is calculated to be 99.991\% at 1064~nm for temperatures below 20~K and 99.976\% at 300~K.

Fabrication of the resonators begins with the deposition of a SiN$_{x}$ hard mask via plasma enhanced chemical vapor deposition. Next, the device geometry is patterned lithographically using a standard positive photoresist. This pattern is then transferred into the SiN$_{x}$ via plasma etching with CF$_{4}$/O$_{2}$. Definition of the resonator geometry in the Al$_{x}$Ga$_{1-x}$As epilayers relies on electron cyclotron resonance etching through the mirror stack using Cl$_{2}$/Ar, with masking provided by the resist/SiN$_{x}$. To undercut the cantilevers, a buffered citric acid solution is utilized~\cite{Kitano1997}. This selective wet etch allows for the removal of the binary GaAs, in this case the substrate, over the low-aluminum content ternary Al$_{0.12}$Ga$_{0.88}$As layers with excellent selectivity~\cite{Huang2007}. During the undercutting process, the SiN$_{x}$ coating protects the top of the mirror surface, while the thick Al$_{0.92}$Ga$_{0.08}$As layer protects the bottom, ensuring minimal surface roughness and maximum reflectivity. To complete the fabrication sequence, the protective SiN$_{x}$ and Al$_{0.92}$Ga$_{0.08}$As layers are removed in a dilute HF solution and the beams are allowed to air-dry after soaking in methanol. The resonators characterized here consist of both fixed-fixed (doubly clamped) and cantilever (singly clamped) beams with a thickness of 5.5~$\mu$m, a nominal width of 50~$\mu$m or 100~$\mu$m, and nominal lengths between 50~$\mu$m and 400~$\mu$m. A scanning electron micrograph highlighting a completed set of cantilevers is shown in Fig.\ 1.

\begin{figure}[htbp]
\centerline{\includegraphics[width=.5\textwidth]{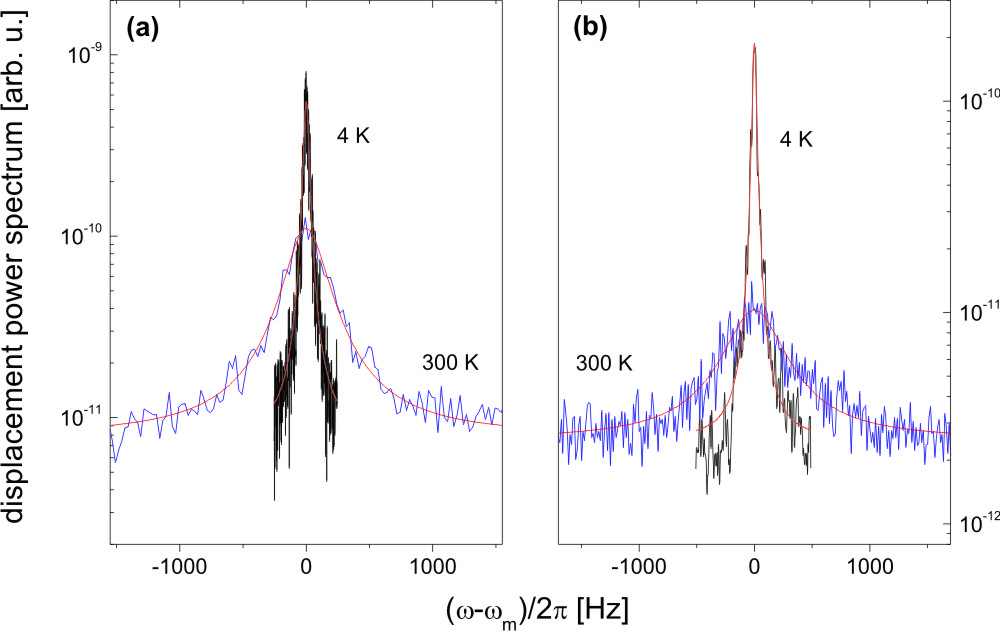}}
\caption{(a) Fundamental mechanical mode of a 150~$\mu$m long doubly clamped resonator at 300~K and 4~K. Central frequencies are 731~kHz and 697~kHz, respectively. The corresponding Q factors are 2200 and 12 000. (b) Second order mode of the same resonator showing Qs of 5000 and 20 000 for frequencies of 1.997~MHz and 1.971~MHz at 300~K and 4~K, respectively.}
\end{figure}

We have characterized the mechanical properties of the resonators optically via interferometric measurements of their displacement. Room-temperature measurements were performed in a standard fiber interferometer~\cite{Rugar1989} while temperature-dependent measurements were carried out using a cryogenic Fabry-P\'{e}rot cavity, in which the micromirror formed one of the cavity's end mirrors (this setup is described in detail in Refs.~\cite{Boehm2006} and~\cite{Groeblacher2008a}). In the case of the fiber interferometer, the displacement power spectrum is directly obtained from the interferometer output, while in the case of the cryogenic Fabry-P\'{e}rot cavity, the noise spectrum of the Pound-Drever-Hall error signal of the cavity is used~\cite{Groeblacher2008a}. At room temperature we obtain mechanical quality factors of up to 7000 for singly clamped and 5000 for doubly-clamped beams. We observe fundamental resonance frequencies of the beams up to 1~MHz in accordance with theoretical estimates based on standard beam theory (see for example, Ref.~\cite{Rao1990}). In particular, we identified a doubly clamped resonator (150$\times$50~$\mu$m) with a fundamental frequency of 730~kHz and higher order resonance at 1.99~MHz. At low temperatures, i.e. operating inside a 4~K helium cryostat, we measure a quality factor of the high frequency mode of 20 000, compared to a Q value of 5000 at room temperature. We observe a similar increase of Q for the fundamental mode of the micromirror, namely from 2200 at room temperature to 12 000 at 4~K (see Fig.\ 2). As expected, the frequency of the resonator modes does not change significantly upon cooling. Cryogenic Q-values of a similar range (10 000$<$Q$<$30 000) have previously been reported for micromechanical resonators fabricated in this materials system~\cite{Harris2000,Mohanty2002}; however, these examples exhibited insufficient reflectivity for our application. Although our devices are not optimized for force detection, we have estimated the thermal force noise of the resonators, which provides an upper bound for the achievable resolution~\cite{Stowe1997}. For the vibration mode near 700~kHz (2~MHz), we calculate an approximate force sensitivity of 220~fN/$\sqrt{Hz}$ (24~fN/$\sqrt{Hz}$) at 300~K, decreasing to roughly 20~fN/$\sqrt{Hz}$ (3~fN/$\sqrt{Hz}$) at cryogenic temperatures. These values are on par with previous examples of GaAs-based nanomechanical resonators as presented in~\cite{Tang2002}.

In order to obtain the micromirror reflectivity we measure the finesse of the Fabry-P\'{e}rot cavity (see above), which provides a measure of the overall intensity losses in the cavity. Knowing the independently determined reflectivity of the macroscopic input mirror (R$_{in}$=99.91\%) one hence obtains a lower limit on the reflectivity R$_{micro}$ of the micromirror. The observed finesse of greater than 5500 [Fig.\ 1(c)] yields a reflectivity R$_{micro}\apprge$99.98\%, in good agreement with the expected values from theory. The reflectivity of our Al$_{x}$Ga$_{1-x}$As Bragg mirrors is comparable to that measured in high-finesse semiconductor microcavities~\cite{Stoltz2005}.

We have demonstrated high-performance micromechanical megahertz oscillators based on free-standing monocrystalline Al$_{x}$Ga$_{1-x}$As DBRs. We observe optical reflectivities exceeding 99.98\% combined with mechanical quality factors up to 20 000 at 4~K for mechanical modes as high as 2~MHz. Given the alleviation of mechanical dissipation compared to previous high reflectivity dielectric stacks, this materials system is an interesting candidate for low-noise optical coatings as needed for example for gravitational-wave detection or for high-precision frequency stabilization of lasers as are used for optical frequency standards. The reported performance can readily achieve an optical finesse of up to 30 000, assuming a matched input coupler reflectivity of R$_{micro}$, allowing these micromechanical devices to operate in a regime of mechanical-sideband limited performance as is required to achieve ground state cavity-cooling of mechanical systems. As the microfabrication process does not deteriorate the reflectivity of the coating, higher finesse values should be achievable by further improving the initial DBR quality.\\

This work was performed under the auspices of the U.S. Department of Energy by LLNL under Contract DE-AC52-07NA27344. We acknowledge financial support by the FWF (Projects P19539-N20 and L426-N20) and by the Foundational Questions Institute fqxi.org (Grant RFP1-06-14). We thank A. Jackson of UCSB for the growth of the Bragg mirror. S. G. is recipient of a DOC-fellowship of the Austrian Academy of Sciences and also acknowledges support from the FWF doctoral program Complex Quantum Systems (W1210).

\end{document}